\documentclass[a4paper]{article}
\usepackage[latin1]{inputenc}
\usepackage{bbm,amsmath,amsfonts,latexsym}
\newcommand{\kk}[3]{\ensuremath{[#1,#2]_{\scriptscriptstyle #3}}}
\newcommand{\ak}[3]{\ensuremath{\{#1,#2\}_{\scriptscriptstyle #3}}}
\newcommand{\je}{\ensuremath{\hat{\mathcal{J}}_{\scriptscriptstyle 1}}}

\newcommand{\tz}{\ensuremath{\hat{\mathcal{T}}_{\scriptscriptstyle 2}}}
\newcommand{\se}{\ensuremath{\hat{\mathcal{S}}_{\scriptscriptstyle 1}}}
\newcommand{\sz}{\ensuremath{\hat{\mathcal{S}}_{\scriptscriptstyle 2}}}
\newcommand{\boe}{\ensuremath{\hat{\mathcal{B}}_0^{(1)}}}
\newcommand{\bod}{\ensuremath{\hat{\mathcal{B}}_0^{(3)}}}

\newcommand{\bozlpe}{\ensuremath{\hat{\mathcal{B}}_0^{(2l+1)}\!\!}}
\newcommand{\cje}{\ensuremath{\mathcal{J}_{\scriptscriptstyle 1}}}

\newcommand{\ctz}{\ensuremath{\mathcal{T}_{\scriptscriptstyle 2}}}
\newcommand{\cse}{\ensuremath{\mathcal{S}_{\scriptscriptstyle 1}}}
\newcommand{\csz}{\ensuremath{\mathcal{S}_{\scriptscriptstyle 2}}}

\newcommand{\cbozlpe}{\ensuremath{{\mathcal{B}}_0^{(2l+1)}\!\!}}
\newcommand{\qje}{\ensuremath{{\hat{\mathcal{J}}\!\!\!\!\backslash}_{\scriptscriptstyle 1}}}

\newcommand{\qtz}{\ensuremath{{\hat{\mathcal{T}}\!\!\!\!\backslash}_{\scriptscriptstyle 2}}}
\newcommand{\qse}{\ensuremath{{\hat{\mathcal{S}}\!\!\!\backslash}_{\scriptscriptstyle 1}}}
\newcommand{\qsz}{\ensuremath{{\hat{\mathcal{S}}\!\!\!\backslash}_{\scriptscriptstyle 2}}}
\newcommand{\qboe}{\ensuremath{{\hat{\mathcal{B}}\!\!\!\backslash}_0^{(1)}}}
\newcommand{\qbod}{\ensuremath{{\hat{\mathcal{B}}\!\!\!\backslash}_0^{(3)}}}

\newcommand{\qbozlpe}{%
  \ensuremath{{\hat{\mathcal{B}}\!\!\!\backslash}_0^{(2l+1)}\!\!}%
}

\newcommand{\bruch}[2]{\ensuremath{\textstyle\frac{#1}{#2}}}

\begin{document}

\title{
  \vspace*{-5em}
  \begin{flushright}
      \small
        THEP 02/14\\
        University of Freiburg\\
        11th October 2002\\
        math-ph/0210024
    \end{flushright}
  \vspace{2em}
  The
  Presentation of the Algebra of Observables of the Closed
  Bosonic String in $1+3$ Dimensions:\\
  Calculation up to Order $\hbar^7$}
\author{G. Handrich$^a$,
  C. Pauf\/ler$^a$\thanks{cornelius.paufler@physik.uni-freiburg.de},
  J. B. Tausk$^a$,
  M. Walter$^a$,
  \\[2ex]
  $^a$ Physikalisches Institut\\
  Albert-Ludwigs-Universität\\
  Hermann-Herder-Str. 3\\
  D-79104 Freiburg i. Br.\\
  Germany
}
\date{}
\maketitle
\begin{abstract}
    We proceed with the investigation of a method of
    quantization of the observable sector  of closed bosonic
    strings.
    For the presentation of the quantum
    algebra of observables the construction cycle concerning
    elements of order $\hbar^6$
    has been carried out.
    We have computed the quantum corrections to the
    only generating relation of order $\hbar^6$. This relation is of spin-parity
    $J^P=0^+$.
    We found that the quantum corrections to
    this relation break the semidirect
    splitting of the classical algebra into an abelian, infinitely generated
    subalgebra $\mathfrak a$
    and a non-abelian, finitely generated subalgebra $\mathfrak U$.
    We have established that there are no (``truly independent'')
    generating relations of order $\hbar^7$.
\end{abstract}

\section{Introduction}

Closed bosonic strings can be described by an infinite set of parametrization
invariant generators and their algebra.
On the classical level they form a Poisson
algebra (\cite{Pohlmeyer:1999}).
Without loss of generality
we restrict ourselves to the investigation of the subalgebra of
``right-movers''
$\mathfrak h^-$,
or rather to its quantum counterpart $\hat{\mathfrak
  h}^-$.
Furthermore, we consider the case of {\em massive} strings moving in $1+3$
dimensional Minkowski space.

In the following it is assumed that the reader is familiar with the content of
ref. \cite{Pohlmeyer:1999} (for a recent review,
cf. ref. \cite{MeusburgerRehren:2002}).
In the present letter, we will adhere to the
notation and terminology employed there.

$\hat{\mathfrak h}^-$ carries an
$\mathbbm N$-gradation in the $\frac{\hbar}{2\pi\alpha'}$-scaling
dimension of its elements ($\alpha'$ denoting the inverse string
tension), for short in ``powers of $\hbar$''. Moreover,
$\hat{\mathfrak h}^-$ carries
a representation of the $O(3)$ symmetry of the target
space. Hence, its elements can be organized in powers of
$\hbar$ and in $so(3)$ spin-parity multiplets, denoted by $J^P$.
The spin of a given multiplet in the algebra will be indicated by a
subscript. In particular, commutator ($[\cdot,\cdot]$) and
anticommutator ($\{\cdot,\cdot\}$) operations can be
organized to yield $so(3)$ multiplets again: let $[\cdot,\cdot]_j$ and
$\{\cdot,\cdot\}_j$
be the respective spin $j$ multiplets.

In ref. \cite{Pohlmeyer:1999}, an algorithm has been given for the
construction of a presentation both of the classical and of the
quantum algebras, starting from the spin operator $\je$ of order
$\hbar$ and the elements \se, \sz, \tz\ and \boe of
power $\hbar^2$ and subsequently
increasing cycle by cycle the $\hbar$-power by
one, introducing new generators \bozlpe\ of the subalgebra
$\hat{\mathfrak a}$ in every even order
$\hbar^{2l+2}$, $l=(0),1,2,\ldots$.
Recall from (\cite[p. 25]{Pohlmeyer:1999})
that \je, \se, \sz, \tz\ and \bozlpe\ are of spin-parity
$1^+$, $1^-$, $2^-$, $2^+$ and $0^+$, respectively.

The consistency of the results of the previous cycles of this
algorithm,
in particular their agreement with
the guiding principle of structural correspondence between the
classical and the quantum
algebra, has been proved recently by Meusburger and Rehren
(\cite{MeusburgerRehren:2002}).

On the classical level, the set of elements \cje, \cse, \csz, \ctz\
generates a Poisson subalgebra
$\mathfrak U$ of ${\mathfrak h}^-$, while the infinite set of elements
\cbozlpe, $l=0,1,\ldots$, forms an abelian subalgebra $\mathfrak a$ of
${\mathfrak h}^-$. Moreover,
it has been observed previously
that the two Poisson subalgebras $\mathfrak a$ and $\mathfrak U$ of
$\mathfrak h^-$, at least up to a ``grade'' (\cite[p. 27]{Pohlmeyer:1999})
which would correspond to
order $\hbar^8$ in the respective quantum algebras, are disjoint and
that the Poisson bracket action of the first two elements of
$\mathfrak a$, \boe\ and \bod, on the elements of $\mathfrak U$ is
semidirect. Among other things our analysis below will clarify the
relevance of this observation for the quantum algebras, with
$\hat{\mathfrak a}$ and
$\hat{\mathfrak U}$ denoting the respective quantum counterparts.

\section{Presentation of the results}

We have computed the cycle of ``degree $5$'' (\cite[pp. 39]{Pohlmeyer:1999})
concerning all the relations in $\hat{\mathfrak U}$ of
order $\hbar^6$.

For this, the only new generating relation of order $\hbar^6$
had to be taken into
account. The classical version of this relation has been calculated by
one of us (G.H.), after its existence had been established in the
course of the
computation of the {\em quantum} action of the generator \bod\ on
the elements of $\hat{\mathfrak U}$ (\cite{HandrichNowak:1999}).
The complete quantum relation
is given in the appendix.
The fact that this relation appears to be a rather long expression
depends crucially on the basis chosen\footnote{
  In our algorithm, this basis is determined by the order in which the
  elements of a given power in $\hbar$ are to be arranged.
  For instance, for pure commutator basis elements, apart from the general
  techniques developed by M. Hall (\cite{Hall:1950}) for the
  presentation of free Lie algebras,
  we have, roughly speaking, chosen the following scheme.
  (The treatment of pure
  anticommutators and mixtures of commutators and anticommutators
  proceeds similarly -- it can be deduced from the
  arrangement of the terms in the $\hbar^6$, $0^+$ relation.)
  First, sort
  the pure commutator terms by the number of
  times each of the generators \tz, \sz, \se\ occurs separately
  in each term. Introduce
  the ordering  $\tz>\sz>\se$.
  Arrange the expressions with equal numbers of occuring \tz, \sz, \se\
  according to the length of the longest uninterupted iterated commutator,
  i.e. of the longest sequence of the form (we suppress the spin subscripts)
  $[\cdots[[g,g'],g''],\cdots,g^{(n)}]$, where the $g^{(i)}$ denote
  generators.
  Then, arrange all pure commutators according to the following rules
  (let $f$ and $g$ be pure commutators):
  $[f,g]_l>[f',g']_{l'}$ if $f>f'$ or if $f=f'$ and $g>g'$,
  or if $f=f'$, $g=g'$ and $l>l'$.
}
in order $\hbar^3$ and higher. Its existence and content,
however, are independent
of the choice of such a basis.

For the definition of the quantum relation, the most general correction
term has been added, introducing 14 parameters of order $\hbar$, 12
of order $\hbar^2$ and 2  of order $\hbar^4$ [sic!]. Note that there
are no elements of the algebra $\hat{\mathfrak h}^-$ that are of order $\hbar^3$ and
spin-parity $0^+$ (\cite[p. 27]{Pohlmeyer:1999});
hence,
there cannot be any $\hbar^3$ correction terms. Among the added
terms which are linearly independent of each other,
there are exactly two, \bod and \boe, from the infinitely
generated abelian
subalgebra $\hat{\mathfrak a}$. Their pertinent coefficients will be denoted
by $x_1$ and $x_2$. They are of respective order $\hbar^2$ and
$\hbar^4$. At first sight, two more $\hbar^4$ correction terms (other
than \bod) seem to
be allowed, to wit the anticommutators $\{\{\je,\je\}_0,\boe\}_0$ and
$\{\boe,\boe\}_0$. However, a computation reveals that e.g. their
commutators with \se\ yield two polynomially independent elements not contained in
$\hat{\mathfrak U}$ which cannot be compensated by other
commutators. On the other hand, by the already established correspondence
between classical
and quantum algebras in the subleading orders of $\hbar$,
there is no room for additional relations (below
$\hbar^6$). Therefore, the respective coefficients of the
above mentioned terms must vanish in the quantum relation.

Following the construction algorithm with the help of computer-algebraic
software\footnote{
  We are using Mathematica routines that have been
  developed by C. Nowak and G. Handrich
  in the course of their lower $\hbar$-order investigation
  (\cite{HandrichNowak:1999}).
  In order to cope with the sheer numbers of equations and
  variables occurring at the present $\hbar$-order --
  in some sectors, systems of as many as $10^4$ equations
  in around $10^4$ variables arise -- we have translated
  part of the routines into the symbolic manipulation
  language FORM \cite{Vermaseren:2000nd}.}
we have obtained
the following results:
\begin{enumerate}
\item
The cycle of induction into the $J^P=0^+$, $\hbar^7$-sector
shows that vanishing coefficients have to
be assigned to all of the $14$ correction terms of order
$\hbar$, in accordance with
the $\mathbbm Z_2$-grading
of the quantum corrections that has been observed in
ref. \cite[p. 48]{Pohlmeyer:1999}.

\item
All of the relations of order $\hbar^7$ can be obtained by
the induction of relations of order $\hbar^4$ and lower. This
observation is based on the fact that
the number of independent elements of a
given $\hbar$-power and spin-parity is known explicitly and the fact
that there do not exist generating relations of order $\hbar^5$.
In other
words, granting the knowledge of the \boe- and \bod-actions on \se, \sz, \tz\
(the former being given in ref. \cite{Pohlmeyer:1999}, the latter in
ref. \cite{HandrichNowak:1999}),
the set of relations given in ref.\cite{Pohlmeyer:1999}
plus the sole generating relation of order $\hbar^6$ gives a complete
description of the finitely generated $\hat{\mathfrak U}$-part of the
algebra of observables at least up to order $\hbar^7$.

\item
Regarding the requirement of correspondence, no restrictions
on
the (as yet undetermined) parameters $f$, $g_1$, $g_2$ have been
found. (These parameters have been introduced in the course of
definition of quantum relations of lower power of $\hbar$
in ref. \cite[p. 44]{Pohlmeyer:1999} and
ref. \cite{HandrichNowak:1999}. Should their values ever be
determined in cycles of
higher power of $\hbar$,
they will assume rational values, cf. ref. \cite[p. 47]{Pohlmeyer:1999}.)
However, all of the $12$ surviving
$\hat{\mathfrak U}$-contributions to the
quantum correction of the $\hbar^6$, $0^+$ relation
can be expressed in terms of $f$ and $g_2$. More precisely,
the eleven coefficients of order $\hbar^2$ are linear functions of $f$,
and the remaining $\hbar^4$ coefficient
is quadratic in $f$ and linear in
$g_2$.

\item
The parameter $x_1$ that describes the contribution of \bod\
in the quantum relation
turns out to be {\em nonzero}.
The numerical value for $x_1$ can be inferred from the quantum relation
given in the appendix.
The fact that $x_1$ is nonzero
means that the semidirect decomposition of the classical algebra
$\mathfrak h^-$ into a sum of the
abelian subalgebra $\mathfrak a$ and the finitely generated
subalgebra $\mathfrak U$ {\sl cannot} be carried over to the quantum
case.

At hindsight, one could have anticipated that elements from the
abelian subalgebra $\hat{\mathfrak a}$ had to be included in the quantum
corrections to the new classical generating relation:
observe that the generating relation of order $\hbar^6$ is the first
encounter with such a relation of characteristic
$J^P=0^+$ and that, in addition,
it is of {\em even} order in powers of $\hbar$, invoking, in
accordance with the $\mathbbm Z_2$-grading, $0^+$ quantum corrections
of {\em even} orders in powers of $\hbar$. The generators of the
abelian subalgebra, \bozlpe, $l=0,1,2,\ldots$
are all of this type. Thus they must be considered as candidates for
the quantum corrections. Moreover,
for every generator\footnote{In this argument, we disregard the generator
  \boe\ because it commutes with the \bozlpe-quantum
  corrections by construction.} \se, \sz, \tz\
there is exactly one way
to induce such a relation together with its most general quantum
correction into
$J^P=1^-,2^-,2^+$ relations of $\hbar$-order increased by
one, respectively. Further, for the present case,
notice that there are only three Lie-type
basis elements (i.e. basis elements that can be expressed as pure
commutators of the generators)
contained in the vector space of characteristic
$\hbar^4$, $J^P=0^+$, not belonging to the subalgebra $\hat{\mathfrak a}$.
Hence, upon the above inductions -- apart from the
generator \bod\ of the abelian algebra $\hat{\mathfrak a}$ -- only these
three elements give rise to linear combinations of
Lie-type basis elements of order $\hbar^5$ (besides elements not of
Lie-type). In other words, at most {\em three}
linearly
independent such combinations contained in
the pertinent vector spaces can be obtained in turn by this procedure.
However, this number is rather poor in comparison with the numbers
$16$, $15$ and $14$
of Lie-type elements of the respective vector spaces. Small wonder that
assistance by the generator $\bod$ of the abelian subalgebra is badly
needed as an additional quantum correction, which, upon the inductions
by \se, \sz\ and  \tz,
each time gives rise to one additional independent
linear combination of Lie-type basis elements of order $\hbar^5$, among
other things.

Because of the ambiguity in the definition of the
subalgebra $\hat{\mathfrak a}$
(\cite{PohlmeyerRehren:1988}, see also
ref. \cite{MeusburgerRehren:2002}), one might hope that by a different
choice for the generators \bozlpe\ of $\hat{\mathfrak a}$
the (quantum) breaking of the semidirect splitting (of the classical
algebra $\mathfrak h^-$) could be avoided.
This is in fact impossible, since already for the present case
every different choice
for the generator \bod\ in the $\hbar^4$-power quantum corrections to
the generating $\hbar^6$-$0^+$-relation results in a mere redefinition
of some of the coefficients of the $\hat{\mathfrak U}$-terms in this part of the relation.

\item
The other parameter, $x_2$, characterizing the \boe\ dependence, is given
as a function linear both in $f$ and in $g_1$.

Using a {\em concrete} realization of the algebra of observables,
Meusburger and Rehren found the variables $f$, $g_1$ and $g_2$ to take
on certain rational numbers, the latter property in agreement with
the results of ref. \cite{Pohlmeyer:1999}. Insertion of these values
into the formula
for $x_2$ reveals a {\em non-vanishing} contribution of \boe\ to the quantum
corrections as well.
\end{enumerate}

A conservative a priori estimate would suggest that the parameters of
the quantum corrections would be roughly tenfold overdetermined. As
yet, the reason for the fact that they are {\em not} overdetermined has not
been identified in terms of special properties of the generating
relations and the actions of \bozlpe,
$l=0,1,2,\ldots$.
Instead, Meusburger and Rehren (\cite{MeusburgerRehren:2002}) give a
reason in terms of the properties of an embedding algebra.

\section*{Acknowledgments}

This project has been carried out under the supervision of K. Pohlmeyer. The
authors wish to thank him
for thorough discussions on the interpretation of the results
and for many suggestions concerning the manuscript. Also, we would
like to thank D. Giulini for a critical reading of the final version
of this letter. This work was supported in part by the DFG-Forschergruppe
Quantenfeldtheorie, Computeralgebra und Monte-Carlo Simulation.

\newpage
\begin{appendix}
\section{The defining relation of order $\hbar^6$}

In order to make the elements of the quantum algebra of observables
and their relations
scale invariant, and in order to make contact with
refs. \cite{Pohlmeyer:1999} and \cite{HandrichNowak:1999} easier,
the generators \je, \se, \sz, \tz\ have been rescaled
according to
\begin{gather*}
    \je=\bruch{\hbar}{2\pi\alpha'}\qje,
    \\
    \se=\left(\bruch{\hbar}{2\pi\alpha'}\right)^2\qse,\quad
    \sz=\left(\bruch{\hbar}{2\pi\alpha'}\right)^2\qsz,\quad
    \tz=\left(\bruch{\hbar}{2\pi\alpha'}\right)^2\qtz
    \\
    \bozlpe=\left(\bruch{\hbar}{2\pi\alpha'}\right)^{2(l+1)}\qbozlpe.
\end{gather*}
The defining quantum relation with its quantum corrections fixed by
consistency is given by the following expression. The rational valued
parameters $f$,
$g_1$ and $g_2$ remain undetermined.
\begin{equation*}
    \label{eq:rel500}
    \begin{split}
        &0=\\
        &-\bruch{1\,729\,i}{5}\, \sqrt{2}\,
        \kk{\kk{\kk{\qtz}
                   {\qse}
                   {2}}
               { \qse}
               { 1}}
           {\kk{\qsz}
               {\qse}
               { 1}}
           { 0}
        -
        {105\,i}\,\sqrt{2}\,
        \kk{\kk{\kk{\qtz}
                   {\qse}
                   {1}}
               {\qse}
               {2}}
           {\kk{\qsz}
               {\qse}
               {2}}
           {0}
        \\&
        +
        \bruch{364\,i}{5}\,\sqrt{6}\,
        \kk{\kk{\kk{\qtz}
                   {\qse}
                   {1}}
               {\qse}
               {1}}
           {\kk{\qsz}{\qse}{1}}
           {0}
        -
        {808\,i}\,{\textstyle\sqrt{\bruch{7}{3}}}\,
        \kk{\kk{\kk{\qsz}{\qse}{3}}{\qse}{3}}{\kk{\qtz}{\qse}{3}}{0}
        \\&
        +
        \bruch{1217\,i}{5}\,\sqrt{\bruch{7}{6}}\,
        \kk{\kk{\kk{\qsz}
                   {\qse}
                   {3}}
               {\qse}
               {2}}
           {\kk{\qtz}
               {\qse}
               {2}}
           {0}
        +
        {881\,i}\,\sqrt{\bruch{14}{3}}\,
        \kk{\kk{\kk{\qsz}{\qse}{2}}{\qse}{3}}{\kk{\qtz}{\qse}{3}}{0}
        \\&
        +
        {259\,i}\,\sqrt{\bruch{1}{30}}\,
        \kk{\kk{\kk{\qsz}{\qse}{2}}
               {\qse}
               {2}}
           {\kk{\qtz}{\qse}{2}}
           {0}
        -
        {105\,i}\,\sqrt{\bruch{1}{2}}\,
        \kk{\kk{\kk{\qsz}{\qse}{2}}{\qse}{1}}
           {\kk{\qtz}{\qse}{1}}
           {0}
        \\&
        +
        \bruch{567\,i}{5}\,\sqrt{\bruch{1}{2}}\,
        \kk{\kk{\kk{\qsz}{\qse}{1}}
               {\qse}
               {2}}
           {\kk{\qtz}
               {\qse}
               {2}}
           {0}
        -
        {273\,i}\,\sqrt{\bruch{3}{2}}\,
        \kk{\kk{\kk{\qsz}{\qse}{1}}{\qse}{1}}{\kk{\qtz}{\qse}{1}}{0}
        \\&
        -
        424\,\sqrt{\bruch{1}{15}}\,
        \kk{\kk{\kk{\kk{\qtz}{\qse}{2}}{\qse}{1}}{\qse}{1}}{\qse}{0}
        +
        {170}\,\sqrt{3}
        \kk{\kk{\kk{\kk{\qtz}{\qse}{1}}{\qse}{2}}{\qse}{1}}{\qse}{0}
        \\&
        +
        {214}\,\sqrt{\bruch{1}{5}}
        \kk{\kk{\kk{\kk{\qtz}{\qse}{1}}{\qse}{1}}{\qse}{1}}{\qse}{0}
        -
        53\,\sqrt{15}
        \kk{\kk{\kk{\kk{\qtz}{\qse}{1}}{\qse}{0}}{\qse}{1}}{\qse}{0}
        \\&
        +
        {25\,200}\,i\,\sqrt{5}\,
        \ak{\kk{\kk{\kk{\qsz}{\qse}{1}}{\qtz}{1}}{\qtz}{1}}{\qje}{0}
        +
        {110\,315}\,\sqrt{\bruch{2}{21}}\,
        \ak{\kk{\kk{\kk{\qtz}{\qse}{2}}{\qse}{3}}{\qtz}{1}}{\qje}{0}
        \\&
        -
        {372\,614}\,\sqrt{\bruch{2}{15}}\,
        \ak{\kk{\kk{\kk{\qtz}{\qse}{2}}{\qse}{2}}{\qtz}{1}}{\qje}{0}
        +
        \bruch{854\,096}{5}\,\sqrt{2}\,
        \ak{\kk{\kk{\kk{\qtz}{\qse}{2}}{\qse}{1}}{\qtz}{1}}{\qje}{0}
        \\&
        -
        {30\,876}\,\sqrt{2}\,
        \ak{\kk{\kk{\kk{\qtz}{\qse}{1}}{\qse}{2}}{\qtz}{1}}{\qje}{0}
        +
        \bruch{11\,074}{5}\,\sqrt{6}\,
        \ak{\kk{\kk{\kk{\qtz}{\qse}{1}}{\qse}{1}}{\qtz}{1}}{\qje}{0}
        \\&
        -
        \bruch{10\,931\,782}{5}\,\sqrt{\bruch{6}{7}}\,
        \ak{\kk{\kk{\qtz}{\qse}{3}}{\kk{\qtz}{\qse}{2}}{1}}{\qje}{0}
        \\&
        +
        {3\,277\,904}\,\sqrt{\bruch{2}{15}}\,
        \ak{\kk{\kk{\qtz}{\qse}{2}}{\kk{\qtz}{\qse}{2}}{1}}{\qje}{0}
        \\&
        +
        \bruch{6\,559\,548}{5}\,\sqrt{2}\,
        \ak{\kk{\kk{\qtz}{\qse}{2}}{\kk{\qtz}{\qse}{1}}{1}}{\qje}{0}
        \\&
        -
        {656\,256}\,\sqrt{6}\,
        \ak{\kk{\kk{\qtz}{\qse}{1}}{\kk{\qtz}{\qse}{1}}{1}}{\qje}{0}
        -
        \bruch{29\,176}{9}\,\sqrt{\bruch{2}{3}}\,
        \ak{\kk{\kk{\qsz}{\qse}{1}}{\kk{\qsz}{\qse}{1}}{1}}{\qje}{0}
        \\&
        -
        \bruch{32\,359\,359\,i}{35}\,\sqrt{\bruch{1}{5}}\,
        \ak{\kk{\kk{\kk{\qsz}{\qse}{2}}{\qse}{2}}{\qse}{1}}{\qje}{0}
        \\&
        +
        \bruch{63\,250\,109\,i}{14}\,\sqrt{\bruch{3}{5}}\,
        \ak{\kk{\kk{\kk{\qsz}{\qse}{2}}{\qse}{1}}{\qse}{1}}{\qje}{0}
        \\&
        -
        \bruch{825\,828\,733\,i}{189}\,\sqrt{\bruch{1}{3}}\,
        \ak{\kk{\kk{\kk{\qsz}{\qse}{1}}{\qse}{2}}{\qse}{1}}{\qje}{0}
        \\&
        -
        \bruch{2\,890\,226\,561\,i}{378}\,\sqrt{\bruch{1}{5}}\,
        \ak{\kk{\kk{\kk{\qsz}{\qse}{1}}{\qse}{1}}{\qse}{1}}{\qje}{0}
        \\&
        +
        \bruch{23\,594\,413\,118\,i}{4\,725}\,\sqrt{\bruch{1}{15}}\,
        \ak{\kk{\kk{\kk{\qsz}{\qse}{1}}{\qse}{0}}{\qse}{1}}{\qje}{0}
     \end{split}
\end{equation*}
\begin{equation*}
    \begin{split}
        &
        -
        \bruch{423\,738\,593}{1\,350}\,\sqrt{\bruch16}\,
        \ak{\kk{\kk{\kk{\qse}{\qse}{1}}{\qse}{1}}{\qse}{1}}{\qje}{0}
        -
        {10\,882}\,\sqrt{\bruch17}\,
        \ak{\kk{\kk{\qtz}{\qse}{3}}{\qtz}{1}}{\qse}{0}
        \\
        &
        +
        {17\,848\,i}\,\sqrt{\bruch{14}{5}}\,
        \ak{\kk{\kk{\qtz}{\qse}{2}}{\qtz}{2}}{\qsz}{0}
        +
        {3\,792}\,\sqrt{\bruch15}\,
        \ak{\kk{\kk{\qtz}{\qse}{2}}{\qtz}{1}}{\qse}{0}
        \\
        &
        -
        {25\,928\,i}\,\sqrt{2}\,
        \ak{\kk{\kk{\qtz}{\qse}{1}}{\qtz}{2}}{\qsz}{0}
        -
        {1\,212}\,\sqrt{3}\,
        \ak{\kk{\kk{\qtz}{\qse}{1}}{\qtz}{1}}{\qse}{0}
        \\&
        -
        \bruch{98\,796\,i}{5}\,\sqrt{\bruch{2}{35}}\,
        \ak{\kk{\kk{\qsz}{\qse}{2}}{\qtz}{2}}{\qtz}{0}
        +
        \bruch{27\,896\,i}{5}\,\sqrt{2}\,
        \ak{\kk{\kk{\qsz}{\qse}{1}}{\qtz}{2}}{\qtz}{0}
        \\&
        +
        \bruch{190\,378}{75}\,\sqrt{\bruch{1}{7}}\,
        \ak{\kk{\kk{\qtz}{\qse}{3}}{\qse}{2}}{\qtz}{0}
        +
        \bruch{34\,592}{15}\,\sqrt{\bruch{1}{5}}\,
        \ak{\kk{\kk{\qtz}{\qse}{2}}{\qse}{2}}{\qtz}{0}
        \\&
        -
        \bruch{82\,412}{25}\,\sqrt{3}\,
        \ak{\kk{\kk{\qtz}{\qse}{1}}{\qse}{2}}{\qtz}{0}
        +
        {9\,492}\,\sqrt{7}\,
        \ak{\kk{\kk{\qsz}{\qse}{3}}{\qse}{2}}{\qsz}{0}
        \\&
        -
        {3\,836}\,\sqrt{\bruch{1}{5}}\,
        \ak{\kk{\kk{\qsz}{\qse}{2}}{\qse}{2}}{\qsz}{0}
        +
        \bruch{6\,883\,i}{5}\,\sqrt{\bruch{6}{5}}\,
        \ak{\kk{\kk{\qsz}{\qse}{2}}{\qse}{1}}{\qse}{0}
        \\&
        -
        \bruch{336}{5}\,\sqrt{3}\,
        \ak{\kk{\kk{\qsz}{\qse}{1}}{\qse}{2}}{\qsz}{0}
        +
        {1\,439\,i}\,\sqrt{\bruch{2}{5}}\,
        \ak{\kk{\kk{\qsz}{\qse}{1}}{\qse}{1}}{\qse}{0}
        \\&
        +
        \bruch{3\,539}{5}\,\sqrt{\bruch{1}{3}}\,
        \ak{\kk{\kk{\qse}{\qse}{1}}{\qse}{1}}{\qse}{0}
        +
        \bruch{4\,186\,i}{5}\,\sqrt{2}\,
        \ak{\kk{\qtz}{\qtz}{1}}{\kk{\qsz}{\qse}{1}}{0}
        \\&
        +
        {8}\,\sqrt{\bruch{1}{15}}\,
        \ak{\kk{\qtz}{\qtz}{1}}{\kk{\qse}{\qse}{1}}{0}
        -
        \bruch{4\,648}{5}\,\sqrt3\,
        \ak{\kk{\qtz}{\qsz}{1}}{\kk{\qtz}{\qsz}{1}}{0}
        \\&
        -
        \bruch{3\,144\,i}{5}\,\sqrt{2}\,
        \ak{\kk{\qtz}{\qsz}{1}}{\kk{\qtz}{\qse}{1}}{0}
        -
        \bruch{6\,888}{5}\,
        \ak{\kk{\qtz}{\qsz}{0}}{\kk{\qtz}{\qsz}{0}}{0}
        \\&
        -
        {3\,522}\,\sqrt7\,
        \ak{\kk{\qtz}{\qse}{3}}{\kk{\qtz}{\qse}{3}}{0}
        -
        \bruch{15\,128}{3}\,\sqrt{\bruch{1}{5}}\,
        \ak{\kk{\qtz}{\qse}{2}}{\kk{\qtz}{\qse}{2}}{0}
        \\&
        -
        {1\,392}\,\sqrt{3}\,
        \ak{\kk{\qtz}{\qse}{1}}{\kk{\qtz}{\qse}{1}}{0}
        +
        {784}\,\sqrt{7}\,
        \ak{\kk{\qsz}{\qse}{3}}{\kk{\qsz}{\qse}{3}}{0}
        \\&
        +
        {1\,708}\,\sqrt{5}\,
        \ak{\kk{\qsz}{\qse}{2}}{\kk{\qsz}{\qse}{2}}{0}
        -
        \bruch{6\,076}{5}\,\sqrt3\,
        \ak{\kk{\qsz}{\qse}{1}}{\kk{\qsz}{\qse}{1}}{0}
       \\&
        -
        {1\,841\,i}\,\sqrt{\bruch{2}{5}}\,
        \ak{\kk{\qsz}{\qse}{1}}{\kk{\qse}{\qse}{1}}{0}
        +
        \bruch{7\,188\,112}{5}\,\sqrt3\,
        \ak{\kk{\kk{\qtz}{\qtz}{1}}{\qtz}{2}}{\ak{\qje}{\qje}{2}}{0}
        \\&
        -
        \bruch{67\,419\,551\,597\,i}{14\,175}\,\sqrt{\bruch{2}{5}}\,
        \ak{\kk{\kk{\qtz}{\qse}{2}}{\qsz}{0}}
           {\ak{\qje}{\qje}{0}}
           {0}
        \\&
        -
        \bruch{233\,715\,466\,133\,i}{9\,450}\sqrt{\bruch{1}{70}}\,
        \ak{\kk{\kk{\qsz}{\qse}{2}}{\qtz}{2}}
           {\ak{\qje}{\qje}{2}}
           {0}
        \\&
        -
        \bruch{531\,966\,503\,i}{2\,025}\,\sqrt{\bruch{2}{5}}\,
        \ak{\kk{\kk{\qsz}{\qse}{2}}{\qtz}{0}}
           {\ak{\qje}{\qje}{0}}
           {0}
        \\&
        +
        \bruch{514\,231\,474\,987\,i}{85\,050}\,\sqrt{\bruch{1}{2}}\,
        \ak{\kk{\kk{\qsz}{\qse}{1}}{\qtz}{2}}
           {\ak{\qje}{\qje}{2}}
           {0}
        \\&
        +
        \bruch{21\,685\,633\,366\,363}{1275750}\,\sqrt{\bruch17}\,
        \ak{\kk{\kk{\qtz}{\qse}{3}}{\qse}{2}}
           {\ak{\qje}{\qje}{2}}
           {0}
        \\&
        +
        \bruch{1\,240\,597\,014\,967}{72\,900}\,\sqrt{\bruch{1}{5}}\,
        \ak{\kk{\kk{\qtz}{\qse}{2}}{\qse}{2}}
           {\ak{\qje}{\qje}{2}}
           {0}
        \\&
        -
        \bruch{168\,553\,073\,449}{283\,500}\,\sqrt{\bruch{1}{3}}\,
        \ak{\kk{\kk{\qtz}{\qse}{1}}{\qse}{2}}{\ak{\qje}{\qje}{2}}{0}
        \\&
        +
        \bruch{236\,150\,183}{27}\,\sqrt{\bruch{1}{15}}\,
        \ak{\kk{\kk{\qtz}{\qse}{1}}{\qse}{0}}{\ak{\qje}{\qje}{0}}{0}
        \\&
        -
        \bruch{906\,152}{5}\,\sqrt{2}\,
        \ak{\ak{\kk{\qtz}{\qtz}{1}}{\qtz}{1}}{\qje}{0}
        +
        \bruch{350\,113\,272}{35}\,\sqrt{2}\,
        \ak{\ak{\kk{\qtz}{\qsz}{1}}{\qsz}{1}}{\qje}{0}
        \\&
        -
        \bruch{87\,423\,547\,i}{189}\,\sqrt{\bruch{5}{3}}\,
        \ak{\ak{\kk{\qtz}{\qsz}{1}}{\qse}{1}}
           {\qje}
           {0}
        -
        \bruch{29\,973\,292\,978\,i}{525}\sqrt{\bruch{1}{15}}\,
        \ak{\ak{\kk{\qtz}{\qsz}{0}}
               {\qse}
               {1}}
           {\qje}
           {0}
        \\&
        -
        \bruch{265\,518\,812\,i}{125}\sqrt{\bruch{1}{7}}\,
        \ak{\ak{\kk{\qtz}{\qse}{3}}{\qsz}{1}}
           {\qje}
           {0}
        -
        \bruch{9\,114\,384\,911\,i}{525}\,\sqrt{\bruch{1}{5}}\,
        \ak{\ak{\kk{\qtz}{\qse}{2}}{\qsz}{1}}{\qje}{0}
        \\&
        +
        \bruch{1\,839\,347\,537}{27}\,\sqrt{\bruch{1}{30}}\,
        \ak{\ak{\kk{\qtz}{\qse}{2}}{\qse}{1}}{\qje}{0}
        +
        \bruch{10\,848\,857\,113\,i}{875}\,\sqrt{3}\,
        \ak{\ak{\kk{\qtz}{\qse}{1}}{\qsz}{1}}{\qje}{0}
        \\&
        -
        \bruch{31\,670\,688}{25}\,\sqrt{\bruch{2}{5}}\,
        \ak{\ak{\kk{\qtz}{\qse}{1}}{\qse}{1}}{\qje}{0}
        -
        \bruch{3\,013\,633\,886\,i}{375}\,\sqrt{\bruch{1}{7}}\,
        \ak{\ak{\kk{\qsz}{\qse}{3}}{\qtz}{1}}
           {\qje}
           {0}
        \\&
        +
        \bruch{20\,403\,199\,969\,i}{525}\,\sqrt{\bruch{1}{5}}\,
        \ak{\ak{\kk{\qsz}{\qse}{2}}{\qtz}{1}}
           {\qje}
           {0}
    \end{split}
\end{equation*}
\begin{equation*}
    \begin{split}
        &
        -
        \bruch{371\,746\,530\,749\,i}{7875}\,\sqrt{\bruch{1}{3}}\,
        \ak{\ak{\kk{\qsz}{\qse}{1}}{\qtz}{1}}
        {\qje}
        {0}
        \\&
        -
        \bruch{234\,320\,627}{27}\,\sqrt{\bruch{1}{10}}\,
        \ak{\ak{\kk{\qse}{\qse}{1}}{\qtz}{1}}{\qje}{0}
        -
        \bruch{682\,624}{5}\,\sqrt{\bruch{3}{35}}\,
        \ak{\ak{\qtz}{\qtz}{2}}{\qtz}{0}
        \\&
        -
        \bruch{767\,912}{5}\,\sqrt{\bruch{3}{35}}\,
        \ak{\qtz}{\ak{\qsz}{\qsz}{2}}{0}
        -
        \bruch{307\,388\,i}{5}\,\sqrt{\bruch{2}{15}}\,
        \ak{\ak{\qtz}{\qsz}{1}}{\qse}{0}
        \\&
        -
        \bruch{106\,292}{5}\,\sqrt{\bruch{1}{5}}\,
        \ak{\qtz}{\ak{\qse}{\qse}{2}}{0}
        -
        \bruch{92\,992\,488}{5}\,\sqrt{\bruch{2}{5}}\,
        \ak{\kk{\qtz}{\qtz}{1}}{\ak{\ak{\qje}{\qje}{0}}{\qje}{1}}{0}
        \\&
        +
        \bruch{5\,630\,367\,098\,038\,i}{127\,575}\,\sqrt{\bruch{1}{7}}\,
        \ak{\kk{\qsz}{\qse}{3}}{\ak{\ak{\qje}{\qje}{2}}{\qje}{3}}{0}
        \\&
        +
        \bruch{2\,636\,102\,359\,228\,i}{14175}\,\sqrt{\bruch{1}{15}}\,
        \ak{\kk{\qsz}{\qse}{1}}
           {\ak{\ak{\qje}{\qje}{0}}{\qje}{1}}
           {0}
        \\&
        +
        \bruch{7\,365\,929\,648}{6075}\,\sqrt{2}\,
        \ak{\kk{\qse}{\qse}{1}}{\ak{\ak{\qje}{\qje}{0}}{\qje}{1}}{0}
        \\&
        -
        \bruch{7\,665\,642\,269\,132}{14\,175}\,\sqrt{\bruch{1}{105}}\,
        \ak{\ak{\qtz}{\qtz}{2}}{\ak{\qje}{\qje}{2}}{0}
        \\&
        +
        \bruch{19\,829\,436\,064}{225}\,\sqrt{\bruch{1}{15}}\,
        \ak{\ak{\qtz}{\qtz}{0}}{\ak{\qje}{\qje}{0}}{0}
        \\&
        +
        \bruch{3\,095\,167\,528\,133}{4\,725}\,\sqrt{\bruch{1}{105}}\,
        \ak{\ak{\qsz}{\qsz}{2}}{\ak{\qje}{\qje}{2}}{0}
        \\&
        -
        \bruch{3\,993\,696\,992}{25}\,\sqrt{\bruch{1}{15}}\,
        \ak{\ak{\qsz}{\qsz}{0}}{\ak{\qje}{\qje}{0}}{0}
        \\&
        +
        \bruch{513\,902\,779\,661\,i}{42\,525}\,\sqrt{\bruch{1}{30}} \,
        \ak{\ak{\qsz}{\qse}{2}}{\ak{\qje}{\qje}{2}}{0}
        \\&
        -
        \bruch{915\,331\,544}{405}\,\sqrt{\bruch{1}{5}}\,
        \ak{\ak{\qse}{\qse}{2}}{\ak{\qje}{\qje}{2}}{0}
        +
        \bruch{8\,485\,612\,798}{2\,025}\,
        \ak{\ak{\qse}{\qse}{0}}{\ak{\qje}{\qje}{0}}{0}\hspace{3em}
        \\&
        -
        \bruch{17\,522\,730\,542\,752}{33\,075}\,\sqrt{\bruch{1}{15}}\,
        \ak{\qtz}{\ak{\ak{\ak{\qje}{\qje}{0}}{\qje}{1}}{\qje}{2}}{0}
        \\&
        +
        \bruch{11\,827\,421\,370\,752}{496\,125}\,\sqrt{\bruch{1}{3}}\,
        \ak{\ak{\ak{\ak{\ak{\qje}
                           {\qje}
                           {0}}
                       {\qje}
                       {1}}
                   {\qje}
                   {0}}
               {\qje}
               {1}}
           {\qje}
           {0}
        \\&-
        (\bruch{1\,854\,884}{5} + {4\,662}f)
        \,i\,\sqrt{\bruch{2}{15}}\,
        \kk{\kk{\qtz}{\qse}{2}}{\qsz}{0}
        \\&
        -
        (\bruch{4\,801\,972}{5} +
        {4\,662}f)\,
        \,i\,\sqrt{\bruch{2}{15}}\,
        \kk{\kk{\qsz}{\qse}{2}}{\qtz}{0}
        -
        (\bruch{1\,727\,516}{5} + \bruch{9\,765}{2}f)\,
        \sqrt{\bruch{1}{5}}\,
        \kk{\kk{\qtz}{\qse}{1}}{\qse}{0}
        \\&
        +
        (\bruch{599\,013\,984}{5} + {8\,389\,263}f)
        \sqrt{\bruch{2}{15}}\,
        \ak{\kk{\qtz}{\qtz}{1}}{\qje}{0}
        \\&
        -
        (\bruch{260\,572\,262\,312}{14\,175}
        + \bruch{931\,520\,317}{45}f)\,i\,\sqrt{\bruch{1}{5}} \,
        \ak{\kk{\qsz}{\qse}{1}}{\qje}{0}
        \\&
        +
        (-\bruch{18\,480\,912\,493\,714}{70\,875}
        + \bruch{1\,771\,689\,913}{450}f)\,\sqrt{\bruch{1}{6}} \,
        \ak{\kk{\qse}{\qse}{1}}{\qje}{0}
        \\&
        -
        (\bruch{42\,282\,912}{25}
        +\bruch{345\,444}{5}f)\,\sqrt{\bruch{1}{5}}\,
        \ak{\qtz}{\qtz}{0}
        -
        (\bruch{4\,908\,848}{5}-{49\,014}f)\,\sqrt{\bruch{1}{5}}\,
        \ak{\qsz}{\qsz}{0}
        \\&
        +
        (\bruch{16\,180\,048}{25}
        - \bruch{38\,094}{5}f)\,
        \sqrt{\bruch{1}{3}}\,
        \ak{\qse}{\qse}{0}
        \\&
        +
        (\bruch{324\,253\,499\,648\,732}{165\,375}
        + \bruch{1\,411\,823\,717\,407}{28\,350}f)
        \sqrt{\bruch{1}{5}}\,
        \ak{\qtz}{\ak{\qje}{\qje}{2}}{0}
        \\&
        + (\bruch{2\,843\,448\,661\,981\,888}{2\,480\,625}
        + \bruch{8\,793\,725\,709\,388}{70\,875}f)
        \,
        \ak{\ak{\ak{\qje}{\qje}{0}}{\qje}{1}}{\qje}{0}
        \\&
        -(\bruch{1\,309\,038\,497\,626\,688}{826\,875}
        +\bruch{35\,892\,576\,098\,324}{23\,625}f
        + \bruch{7\,805\,728\,654}{75}f^2 - {75\,264}g_2)
        \sqrt{\bruch{1}{3}}
        \ak{\qje}{\qje}{0}
        \\&
        -{50\,176}\,\qbod
        +( \bruch{5\,513\,476\,864}{135} + \bruch{2\,747\,136}{5}f
        + {50\,176} g_1)\,\qboe
    \end{split}
\end{equation*}

The last eleven lines constitute the quantum corrections to the
classical relation.
\end{appendix}
\end{document}